\documentclass[prl,twocolumn,showpacs]{revtex4}
\usepackage[english]{babel} \usepackage{amssymb}
\usepackage{amsmath}

\begin{document}
	
\title{Response to the Comment on ``Paradox of Photons\\
Disconnected Trajectories Being Located by Means\\ of `Weak
Measurements' in the Nested Mach-Zehnder\\ Interferometer''
(JETP Letters 105, 152 (2017))}
	
\author{G.\,N. Nikolaev}

\email[e-mail: ]{nikolaev@iae.nsk.su}

\affiliation{Institute of Automation and Electrometry of SB RAS,
Novosibirsk, 630090 Russia\\ Novosibirsk State University,
Novosibirsk, 630090 Russia}

\noindent

\begin{abstract} In [JETP Lett. 105(3), 152 (2017)], a clear and
comprehensive analysis of the paradoxical results of experiment
[Phys. Rev. Lett. 111, 240402 (2013)] was carried out on the
basis of the classical wave theory of light, which presupposes
the continuity of possible of light paths. It was shown that the
paradoxical results of the experiment are due not to the
discontinuity of the trajectories of light, as claimed in [Phys.
Rev. Lett. 111, 240402 (2013)], but to the used way of detecting
the path of photons. The experiment modification proposed in
[JETP Lett. 105(3), 152 (2017)] allows us to eliminate the
seeming discontinuity of the light trajectories. In Comment
[arXiv:1705.02137 (2017)] to the article such modification is
declared unreasonable. This Response to the Comment shows that
this statement is not based on clear and logical arguments.
Instead, it is only asserted that the proposed modification
"violates the faithfulness indication of the trace" of photons.
Therefore, the Comment's criticism can not be considered as
well-founded. Consequently, the conclusion of [JETP Lett.
105(3), 152 (2017)] that a new concept of disconnected
trajectories proposed by the authors of work [Phys. Rev. Lett.
111, 240402 (2013)] is unnecessary, remains valid.
\end{abstract}


\maketitle

The author of the comment \cite{arXiv:quant-ph/1705.02137} on my
work \cite{JETP.Lett.105.475} states that the proposed
modification of the scheme of the experiment reported in
\cite{Phys.Rev.Lett.111.240402} is improper and that the
experiment faithfully demonstrates a discontinuous trace of
photons. It is difficult to agree with this statement because of
the absence of clear and well-founded reasons (see below).

In \cite{JETP.Lett.105.475}, I discussed the recent experiment
\cite{Phys.Rev.Lett.111.240402} with a nested Mach-Zehnder
interferometer. The authors treated the experiment so that the
past of photons is not described by continuous trajectories
\emph{because photons are detected inside the inner Mach-Zehnder
interferometer and are not detected outside it}.

\emph{The idea} of a small modification of the scheme of the
experiment \cite{Phys.Rev.Lett.111.240402} \emph{has been
proposed just to detect photons at the input of the inner
Mach-Zehnder interferometer. Instead of the vibration of the
mirror, the direction of polarization of light is modulated in
this scheme}. Further, in one of the arms of the inner
Mach-Zehnder interferometer, this modulation is converted into a
shift of the light beam of orthogonal polarization, which is
then transformed to the original polarization.

The realization of the idea is as follows. A birefringent plate,
a phase compensator, and two polarizers that oriented at the
angle of $\mathrm{45^{o}}$ and in parallel to the initial
polarization are placed in one of the arms of the inner
Mach-Zehnder interferometer. The exact same polarizers are
placed in the other two arms of the nested Mach-Zehnder
interferometer.

The author of the comment states that such a modification is
improper because it \emph{``violates the faithfulness of the
experiment since it causes different transformations of the
polarization record at E into a signal at the detector,
depending on the path the beam takes''}. The author of the
comment also claims that ``the signal at the detector does not
present \emph{a faithful indication} of the trace inside the
interferometer'' \emph{in the absence of conditions that ``the
shift of the direction in the region of the local coupling is
translated to the shift of the output beam in the same way for
all possible paths of the beam toward the detector''}. It is not
explained what is the ``faithfulness'' of the experiment and
indication and how it is logically related to the cited
conditions. \emph{The necessity of these conditions was not
justified either in \cite{Phys.Rev.Lett.111.240402} or in
subsequent works of these authors
\cite{Phys.Rev.A87.052104,Phys.Rev.A89.024102,fphy.2015.00048,Phys.Rev.A93.17801,
Phys.Rev.A93.036103,arXiv:quant-ph/1610.04781,arXiv:quant-ph/1610.07734,Chin.Phys.Lett.34.020301}}.

The authors' treatment of the two-state vector formalism is
attractively simple but is not justified and, thereby, can
provide false conclusions and is limited. This concerns the
\emph{interpretation} of the results of the experiment as
\emph{the manifestation of the ``discontinuous trace''} of
photons by the authors of \cite{Phys.Rev.Lett.111.240402}.
\emph{This same treatment does not explain the absence of any
signals when the beam of light in the lower arm of the outer
Mach-Zehnder interferometer is blocked}. At the same time, this
absence is clearly explained by the traditional theories of
light and quantum mechanics. \emph{Therefore, it is unnecessary
to introduce a new concept of discontinuity of possible paths of
photons}.

\end{document}